\documentclass[journal abbreviation, journal]{copernicus}

\begin{document}

\title{COSMIC RAY TRANSPORT AND ANISOTROPIES TO HIGH ENERGIES}


\Author[1,2,3,4]{Peter L.}{Biermann}
\Author[5]{Lauren\c{t}iu  I.}{Caramete} 
\Author[6]{Athina}{Meli}  
\Author[7]{Biman N.}{Nath}
\Author[8]{Eun-Suk}{Seo}
\Author[9]{Vitor de}{Souza} 
\Author[10]{Julia Becker}{Tjus}  
\affil[1]{MPI for Radioastronomy, Bonn, Germany}
\affil[2]{Dept. of Phys., Karlsruhe Inst. for Technology KIT, Karlsruhe, Germany}
\affil[3]{Dept. of Phys. \& Astr., Univ. of Alabama, Tuscaloosa, AL, USA}
\affil[4]{Dept. of Phys. \& Astron., Univ. Bonn, Bonn, Germany}
\affil[5]{Institute for Space Sciences, Bucharest, Romania}
\affil[6]{Department of Physics and Astronomy, University of Gent, Gent, Belgium}
\affil[7]{Raman Research Institute, Bangalore, India}
\affil[8]{Dept. of Physics, Univ. of Maryland, College Park, MD, USA}
\affil[9]{Univ. de S${\rm \tilde{a}}$o Paulo, Inst. de F\'{\i}sica de 
S${\rm \tilde{a}}$o Carlos, S${\rm \tilde{a}}$o Carlos, Brazil}
\affil[10]{Dept. of Phys., Univ. Bochum, Bochum, Germany}



\runningtitle{Cosmic ray transport}

\runningauthor{Biermann}

\correspondence{P.L. Biermann (plbiermann@mpifr-bonn.mpg.de)}

\received{}
\revised{}
\accepted{}
\published{}


\firstpage{1}

\maketitle

\begin{abstract}
A model is introduced, in which the irregularity spectrum of the Galactic magnetic field beyond the dissipation length scale is first a Kolmogorov spectrum $k^{-5/3}$ at small scales $\lambda \, = \, 2 \pi/k$ with $k$ the wave-number, then a saturation spectrum $k^{-1}$, and finally a shock-dominated spectrum $k^{-2}$ mostly in the halo/wind outside the Cosmic Ray disk.  In an isotropic approximation such a model is consistent with the Interstellar Medium (ISM) data.  With this model we discuss the Galactic Cosmic Ray (GCR) spectrum, as well as the extragalactic Ultra High Energy Cosmic Rays (UHECRs), their chemical abundances and anisotropies.  UHECRs may include a proton component from many radio galaxies integrated over vast distances, visible already below 3 EeV.
\end{abstract}

\introduction  
The spectrum of Cosmic Rays (CRs) ranges from about a  hundred MeV to a few hundred EeV, with a kink down around a few PeV, the "knee", a second kink up at around 3 EeV,  the "ankle", and a sharp turn-off around 100 EeV (PeV $= \, 10^{15}$ eV, EeV $= \, 10^{18}$ eV).  These particles are injected and accelerated in various sites, usually thought to be exploding stars, such as Supernovae (SNe) or Gamma Ray Bursts (GRBs); cf. \citep{IceCube12} for energies below a few EeV, and possible activity of super-massive black holes (Active Galactic Nuclei or AGN) at the higher energies, e.g.,  \citep{Gregorini84,Biermann87,Witzel88,Meli08}.  At their production particles at all energies show a different spectrum, probably flatter, and get steepened by transport through the scattering medium of a turbulent ionized magnetic gas, as the higher energy particles escape faster in what is usually called a "Leaky Box approximation".  The spectrum of magnetic irregularities can be  directly studied in the Solar wind, where a Kolmogorov law is capable of describing the data \citep{Kolmogorov41a,Kolmogorov41b,Goldstein95}. Turbulent energy is injected at some large length scales, producing a saturation spectrum of $k^{-1}$, down to an outer scale $l_{out}$ of a cascade; from there on down in wavelength (or up in wave-number) the turbulent energy flow in a 3D-approximation is constant, giving $k^{-5/3}$, a Kolmogorov cascade \citep{Kolmogorov41a,Kolmogorov41b}; there may be an intermediate range in wave-numbers between $k^{-1}$ and $k^{-5/3}$.  This energy flow goes down to the dissipation scale.  The data confirm this expectation \citep{Armstrong95,Goldstein95,LOFAR13,Haverkorn13}.  Does this behavior also hold for the spectrum of magnetic irregularities in the ISM, and how this spectrum extends to the halo, where a wind might be sweeping everything away \citep{Everett08,Everett12}?  Does this have consequences for our interpretation of the sky distribution of the incoming UHECRs \citep{Auger15}?

\section{The irregularity spectrum in the ISM: Kolmogorov ?}
The electron density fluctuation spectrum in the ISM matches the expectations from a Kolmogorov description \citep{Armstrong95,Haverkorn13}, with an inner scale between $10^{7}$ cm and $10^{10}$ cm, and an outer scale of order 20 pc, confirmed by LOFAR \citep{LOFAR13}; cf. also Voyager data \citep{Voyager15}.  CR data allow to test whether a Kolmogorov description is adequate:  A Kolmogorov cascade implies that the time-scale for diffusion out of the disk of the Galaxy scales as $E^{-1/3}$:  Nuclei are observed to run approximately as $E^{-8/3}$ \citep{CREAM09}, electrons approximately as $E^{-10/3}$ \citep{AMS14}; electrons are in the loss-limit, i.e. losses due to Synchrotron and Inverse Compton interaction dominate over diffusion \citep{Kardashev62}, so their spectrum is expected to be steepened by unity.  Therefore the original spectrum for both, nuclei as electrons, is deduced to be $E^{-7/3}$, demonstrating consistency, excluding those nuclei, such as Fe, for which spallation losses flatten the spectrum, or that energy range for electrons, where additional processes occur, close to TeV energies.  Error bars in the data do not significantly modify the conclusion.  
The mean free path $\lambda_{mfp}$ for scattering of CRs can be written as
$\lambda_{mfp} \, = \, \frac{L_{out}}{b} \, {(E/\{Z \, e \, B \, L_{out}\})}^{1/3} $, which is in turn
$= \, 10^{17.8} \, {\rm cm} \, b_{1/3}^{-1} \, L_{out, 20 \, {\rm pc}}^{2/3} \, {(E_{4 \, {\rm GeV}}/\{Z B_{5 \, {\rm \mu Gauss}}\})}^{1/3}$
where momentum $p c$ is here approximated as energy $E$ and $Z$ is charge of the CR nucleus considered,  $E_{4 \, {\rm GeV}}$ is that energy in units of 4 GeV, $b_{1/3}$ is the fraction of magnetic field energy density in the irregularities in units of 1/3, $L_{out, 20 \, {\rm pc}}$ is the outer scale of the Kolmogorov range in units of 20 pc, and $B_{5 \, {\rm \mu Gauss}}$ is the magnetic field strength in units of 5 $\mu$Gauss.  The scattering coefficient $\kappa$ can then be written \citep{Drury83} as
$\kappa \, = \, 10^{27.8} \, {\rm cm^2/s} \,  b_{1/3}^{-1} \, L_{out, 20 \, {\rm pc}}^{2/3} \, {(E_{4 \, {\rm GeV}}/\{Z  B_{5 \, {\rm \mu Gauss}}\})}^{1/3} .$   This in turn yields an escape time at 4 GeV for a CR disk thickness $H_{CR, 1.5 \, {\rm kpc}}$ in units of 1.5 kpc of $10^{8.1} \,{\rm yrs} \,   H_{CR, 1.5 \, {\rm kpc}}^2 \, b_{1/3}^{+1} \, L_{out, 20 \, {\rm pc}}^{-2/3} E_{4 \, {\rm GeV}}^{-1/3} Z^{+1/3} B_{5 \, {\rm \mu Gauss}}^{-1/3}$ - too long by an order of magnitude \citep{Brunetti00}.  One obvious solution is to insert another turbulence regime to add energy at a different wave-number range, between the outer scale of Kolmogorov turbulence, and the very maximum scale, the thickness of the CR disk, or even beyond, and so reduce $b$, the strength of the irregularities just in the Kolmogorov range:  This implies that $b$ is then best   $b \, \simeq \, 0.03$.
We note that the lateral radius from where sources contribute is of similar scale as the half width scale height of the CR disk, in the case of isotropic scattering; in case of anisotropic scattering that lateral reach may be larger, reducing any expected anisotropy.
Very often, the transport of CRs is deduced from the amount of material it traversed.  Since all the nuclei of heavy elements among the CRs traverse the dense region around the stellar wind and OB-super-bubble \citep{Binns08,Binns11,Binns13}, the interaction may well be decoupled from the transport.  Using the CRs themselves as source of excitation for the wave-field in which they scatter, yields an expected grammage $X_{lb}$-momentum power-law dependence of exponent $-5/9$, in addition to the grammage experienced in the average ISM with an exponent of $-1/3$ \citep{Biermann98,Biermann01,Biermann09}.  This is fully consistent with the derivation by \citep{Ptuskin99}:  $X_{lb} \, = \, 11.3 \, {\rm g/cm^2} \; \beta$ at $R \, < \, 5$  GV, and 	$X_{lb} \, = \, 11.3 \, {\rm g/cm^2}\; \beta \, {(R/\{{\rm 5 \, GV}\})}^{-0.54}$ at $R \, \geq \, 5$ GV, where $\beta \, c$ is the velocity of the nucleus considered, and $R$ is its rigidity $p c/\{Z \, e \}$.
There are many caveats with using such a simple isotropic 3D law as a Kolmogorov $k^{-5/3}$ law of turbulence in magnetic field fluctuations as well as in density and velocity fluctuations;  1)  Magnetic fields are never really isotropic; 2)  The ISM has several phases, with vastly different temperatures and densities, with the most tenuous medium of a density of order $10^{-2.5}$/cc  \citep{Cox74,LagageCesarsky83,Snowden97};  3)  Transonic flow and shocks may be quite common \citep{Iacobelli14}.  Intermittency may be extreme, stoked by, e.g., stellar explosions, active stars such as micro-quasars \citep{Mirabel10}, HII-regions, OB-super-bubbles, and pulsars;    4)  The fraction of magnetic field energy in irregularities differs between spiral arm and inter-arm regions \citep{Beck96};  5)  The CR layer thickness may vary strongly throughout a galaxy, small in the inner regions \citep{Biermann10}, of perhaps only 100 pc or a small multiple thereof, and large outside \citep{Beuermann85}, of kpc scale;  6)  Galaxies often have winds  \citep{RossaDettmar03,Everett08,Everett12,Uhlig12};  7)  Such winds are likely to be highly non-stationary, since the stellar activity driving them is often characterized by bursts. These winds are also probably highly unstable to clump formation, like the winds of massive stars, as any driving by a "light fluid", such as photons (massive stars) or CRs (galactic winds) is unstable;  8)  Driving the irregularity spectrum occurs on many scales.  There may be a further outer scale, at which the scattering  length becomes independent of energy, or the spectrum runs as $k^{-2}$, possibly driven by shocks.  It is amazing, that a simple picture carries us so far \citep{Yan12}.  Accepting it, we can ask whether the knee in the spectrum of CRs can possibly be caused by a change in propagation:  The knee is at an energy of about 2 $Z$ PeV, and its Larmor radius corresponds to a wavelength right in the simple power-law 3D Kolmogorov spectrum of the irregularity spectrum.  So in this approximation the propagation across the knee does not change, e.g. \citep{Biermann93CRI,Todero15}, and the knee must be a feature of the sources.

\subsection{Local CR gradients}
Moving through an isotropic distribution of CRs in their own frame produces an apparent anisotropy, the Compton-Getting effect, \citep{Compton35}.  The anisotropy is given by $\{\delta F_{CR}\}/F_{CR} \, = \, 4 \beta \cos \theta$, where $\beta \, c$ is the velocity relative to the local system of scattering rest, and $\theta$ is the azimuth relative to the direction of motion. Within the scale-height of about 1 to 2 kpc, within the CR-escape time \citep{Brunetti00} there are about between 1000 and 4000 old Super-Nova Remnants (SNRs) and associated magnetic bubbles, with their shells, and pulsars.  Most SNRs are either in cooling, or even later, in the coasting phase.  Following \citep{LagageCesarsky83} and \citep{Cox72} the final radius of the expanding shell of the cooling phase might reach 300 pc for the density $10^{-2.5}$/cc, or 30 pc for density unity per cc.  So already the injection of CRs may be strongly overlapping \citep{Cox74}, reducing their anisotropy.  A corollary question is whether we see in the data all the mottledness of all the nearby CR injection sites \citep{Biermann13}.

There is another source of information about a possible CR gradient, which would result in an anisotropy:  The star formation rate in the Galaxy strongly depends on radial distance from the Galactic Center (GC) \citep{SmithBM78}.  This star formation and therefore the expected SN-rate strongly peak at the GC.  So we expect both the pion contribution of the gamma-ray emission, as well as the local neutrino emission to also show such a peak.  The neutrino emission, which statistically cannot be distinguished from isotropy at present \citep{IceCube14}, does show a cluster of events near the GC.  Assuming the local spectral shape of the CRs, including the knee of the CRs, to also describe the CR spectrum elsewhere in the Galaxy then predicts that the neutrinos ought to have a kink down around 100 TeV.  This is not seen in the cluster of events near the GC, supporting the doubts about an association of all these events with the GC region.

\subsection{ Local Galactic wind?}
A Galactic wind provides a boundary condition for diffusion of GCR particles:  Rossa \& Dettmar (2003) show that, to cite, "a good correlation exists between the far-infrared flux ratio ($S_{60\mu}/S_{100\mu}$) and the Star Formation Rate (SFR) per unit area ($L_{FIR}/D_{25}^2$), based on the detections/non-detections."  A minimal energy input due to star formation per unit area of {  ${\dot{E}}^{thresh}_{25} \, \simeq \,  10^{40.0 \pm 0.3} \,  \rm {erg \, {s}^{-1} \, {kpc}^{-2}}$} has been derived, to cite:  "Gaseous halos are a direct consequence of Star Formation activity in the underlying galactic disk."  Radio observations also demonstrate that many galaxies have winds \citep{Chyzy06,Chyzy07}.  Local energy injection with $N_{SN, 3}$, the number of local supernovae within about 3 kpc distance within the last 1000 years, is: $N_{SN, 3} \; (\{10^{51} \, {\rm erg}\}/\{10^{3} \, {\rm {yr}} \, \pi 3^2  \, {\rm kpc^{+2}}\})$  \citep{Tuellmann06}; this reaches the threshold given if $N_{SN, 3} \, > \, 5$, exceeded slightly by known historical supernovae.  We interpret this as supporting a local Galactic wind \citep{Everett08,Everett12,Uhlig12,Sarkar15}, with about 1 $M_{\odot}$ falling back per year. Beuermann et al. (1985) have derived a model for the CR disk, showing evidence for a thick CR disk, with a half width of 1 - 2 kpc. We interpret the thick disk as the full disk for the confinement of CRs, which provides the inner boundary layer for the wind.  Bending in a Parker wind \citep{Parker58} can be an order of magnitude larger than a local estimate would suggest due to integrating the Lorentz force with an $1/r$ magnetic field.  A galactic wind scatters incoming UHECR  particles via turbulence with many shocks: an ensuing $k^{-2}$ spectrum in irregularities gives $\kappa_{scatt} \, = \, const(E)$,   so observed UHECR spectra remain unaffected by transport locally.

\subsection{Chemical composition and anisotropy at high energy}
The chemical composition rapidly gets heavier with energy towards a particle energy of order 100 PeV \citep{etTilav13}, then gets quickly lighter again, to be compatible with a majority fraction of Hydrogen around 1 to 3 EeV \citep{Erice14}: R. Engel, only to rise again to a heavier composition at higher energies \citep{Auger14a,Auger14b,Todero15}.  We can interpret these data using Galactic CRs, a mixed component including heavy elements from the radio galaxy Cen A, only {50$^{o}$} from the GC in direction, and in addition, possibly an extragalactic proton component from radio galaxies all around \citep{Todero15}.  The Auger sky suggests a cluster of events around the direction towards Cen A \citep{KoteraOlinto11}, a direction also especially abundant in radio galaxies.  So any anisotropy attributable to the GC region and Cen A both ought to show up in the same hemisphere.  On the other hand, the TA experiment \citep{TA14} detects a hot spot in the sky not very far from the strongest local starburst galaxy M82, suggesting GRBs as a possible source.  At energies below 3 EeV and at higher energies we see a directional preference into the opposite direction, at the percent level \citep{Erice14}: R. Engel; \citep{LetessierSelvon14}:  it may correspond to the protons from radio galaxies all around us \citep{Das08}.  If this were true, then with increasing heavy element abundance in the UHECR data \citep{etTilav13}, this direction might be distinguishable via the chemical composition.  We can interpret the 15 degree scattering cloud around the radio galaxy Cen A in the small angle limit.  Using the scale $15 \, R_{15}$ kpc as the maximum scale (twice the Sun's Galactic orbit), a charge of $ 6 \, Z_{6}$, i.e. Carbon, an energy of $6 \cdot 10^{19} \, E_{60 \, {\rm EeV}} $ eV, and $N$ scatterings we find:  $N \, = \, 10^{3.0} \, R_{15 \, {\rm kpc}}^2 \, B_{5 \, {\rm \mu Gauss}}^{2} \, Z_{6}^{2} \, E_{60 \, {\rm EeV}}^{-2} $ from the size of this scattering cloud.  If we set the scattering length scale again to be of the order of the CR disk scale, we have $N$ of the order 10  to perhaps 100 even in the innermost region of the Galaxy \citep{Biermann10}; it follows that $B$ is of $\mu$Gauss level; assuming the cloud of events to be mostly low $Z$ elements implies a stronger magnetic field.  The fact that the cloud of events is pretty much centered on Cen A also means that the linear bending integrates out to near zero; just the fluctuations produce an angular dispersion.  If we accept the chemical composition derived from Auger data \citep{Gopal10}, then other sources further distant are not viable at very high energy.  With a model for the wind density and Galactic mass loss \citep{Everett12} we can derive further  plausible conditions on the wind parameters, and test them.  Any widespread proton component may require many sources.  Radio galaxies pointed at the observer \citep{Gregorini84,Witzel88} are recognized as all variable bursting sources (see, e.g, Her A, \citep{Gizani03}): Their maximally allowed particle energy can be high \citep{Lovelace76}, and so a sufficient number of sources with the required particle energies may exist.  Just a few nearby strongly fluctuating bursting sources may dominate the sky \citep{Lemoine97}; see TA and Auger \citep{AugerTA15}.

\conclusions  
A model is introduced, in which the irregularity spectrum of the Galactic magnetic field beyond the dissipation length scale is in an isotropic approximation first a Kolmogorov spectrum $k^{-5/3}$, then a saturation spectrum $k^{-1}$, and finally a shock-dominated spectrum $k^{-2}$ perhaps mostly in the halo/wind outside the Cosmic Ray disk.  Such a model is clearly a simple isotropic approximation.  With this model we discuss the Galactic Cosmic Ray spectrum, its anisotropies, as well as the extragalactic ultra high energy Cosmic Rays, their chemical abundances and anisotropies.  We have Galactic Cosmic Rays, a component from the radio galaxy Cen A, in direction not far from the Galactic Center, and in addition, possibly an extragalactic proton component from bursting radio galaxies.



\begin{acknowledgements}
Todor Stanev has contributed to the evolution of these ideas over very many years.  Memberships of some of the authors in the AMS, Auger, H.E.S.S., IceCube, Kaskade-Grande, and LOPES collaborations are noted.  Discussions by PLB with W.R. Binns, A. Chieffi, R. Engel, H. Kang, and D. Ryu are gratefully acknowledged.  PLB also thanks Carola Dobrigkeit Chinellato and Roger Clay for  helpful comments on the manuscript.
\end{acknowledgements}

\end{document}